# AFM study of hydrodynamics in boundary layers around micro- and nanofibers

Julien Dupré de Baubigny,[a,b] Michael Benzaquen,[c,$] Caroline Mortagne,[a,b] Clémence Devailly,[d] Sébastien Kosgodagan Acharige,[d] Justine Laurent,[d] Audrey Steinberger,[d] Jean-Paul Salvetat,[e] Jean-Pierre Aimé,[f] Thierry Ondarçuhu*[a]

[a] Nanosciences group, CEMES-CNRS, UPR 8011, 29 rue Jeanne Marvig, 31055 TOULOUSE

[b] Université de Toulouse, 29 rue Jeanne Marvig, 31055 TOULOUSE

[c] UMR CNRS 7083 Gulliver, ESPCI ParisTech, PSL Research University, 10 rue Vauquelin, 75231 PARIS.

[$] Currently at Capital Fund Management, 23 rue de l'Université, 75007 PARIS

[d] Université de Lyon, ENS de Lyon, CNRS, Laboratoire de Physique, 69342 LYON

[e] CRPP, CNRS UPR 8641, 115 Avenue du Dr Albert Schweitzer, 33600 PESSAC

[f] CBMN, CNRS UMR 5248, 2 rue Escarpit, 33600 PESSAC

## ABSTRACT

The description of hydrodynamic interactions between a particle and the surrounding liquid, down to the nanometer scale, is of primary importance since confined liquids are ubiquitous in many natural and technological situations. In this paper, we combine three non-conventional atomic force microscopes to study hydrodynamics around micro- and nano-cylinders. These complementary methods allow the independent measurement of the added mass and friction terms over a large range of probe sizes, fluid viscosities and solicitation conditions. A theoretical model based on an analytical description of the velocity field around the probe shows that the friction force depends on a unique parameter, the ratio of the probe radius to the thickness of the viscous boundary layer. We demonstrate that the whole range of experimental data can be gathered in a master curve which is well reproduced by the model. This validates the use of these AFM modes for a quantitative study of nano-hydrodynamics, and opens the way to the investigation of other sources of dissipation in simple and complex fluids down to the nanometer scale.

# INTRODUCTION

The design of multiscale functional networks with micro-fluidic channels produces a wealth of new experiments and concepts in which the attempt to understand and control the flow of heterogeneous fluids bearing micro- and nano-particles is at the first place. There are now many possible ways to study properties at a microscopic level [1], giving new insights on phenomena that often exhibit features at the macroscopic scale. Applications are numerous in many transversal domains of major interest where the behavior of confined fluid is of primary importance. Within this framework, determining the relevant lengths and scaling laws that govern hydrodynamic interactions is a major goal. The flow around particles is essential to interpret dynamic light scattering experiments [2] or to understand the rheological properties of colloidal dispersions [3]. In particular, the transport properties of rod-like particles [2] has known a renewed interest due to the recent development of carbon nanotubes suspensions [4]. The control of flow inside channels is also of primary importance for the further development of micro- and nano-fluidics. Indeed, many digital fluidic networks are elaborated for screening purposes where spatially localized chemical reactions are planned to operate as a hierarchically organized set of logical gate functions [5]. In the case of flow of suspensions confined inside micro-channels, the hydrodynamic interactions mediated by the embedding liquid lead to anomalous diffusion of the particles [6]. Confined complex fluids are also ubiquitous in life science since many biological processes involve biofluids inside vascular systems[7] or, at a smaller scale, in aquaporin [8]. A fine understanding of the microscopic hydrodynamic coupling between particles is also crucial for the controlled collective motion [9-11] of assemblies of motile particles such as micro- or nano-swimmers [12].

The hydrodynamic interaction can also be used to manipulate nano-objects as, for example, the flow induced structuring of colloidal suspensions [13] or the translocation and stretching of polymers in nanochannels for ultrafiltration [14]. The stress resulting from the fluid velocity field gradient at the wall can even induce the scission of nanotubes under sonication [15].

In all these systems one has to carefully manage the boundary constraints that determine the fluid mechanical properties and flow behavior. In addition to the interface wall on which the fluid molecules can stick on or slide, geometry and size effects of the system, whether a channel or a particle, also matters.

In the present work, we use three different AFM modes to extract the conservative and dissipative contributions of a small volume of fluid surrounding an oscillating nano-object. The aim is to quantify to which extent the surrounding fluid is perturbed by a nano-particle motion. A quantitative knowledge of the fluid contribution, and of the extension of the velocity field upon the action of a unique moving nano-object, first provides information on the energy one has to supply to ensure a stationary state, and second gives the length that determines the range of the hydrodynamic interactions.

Few techniques are available to probe locally the flow behavior around probes with micro- or nano-meter scale. Microrheology techniques have been developed recently to address some of the above mentioned issues [16, 17]. They usually rely on the monitoring of the Brownian motion of micron sized beads or on the measurement of their interaction with the fluid when manipulated by optical tweezers. The results of these passive or active methods are related through the fluctuation-dissipation theorem. The mechanical response of microfabricated cantilevers immersed into the fluid under study has also been used but is limited to gases [18] or liquids of low viscosities (< 10mPa.s) because of the strong damping of the cantilever oscillation [19, 20]. Interaction with polymer layers has also been studied by such techniques [21]. An alternative is to use a cantilever with a hanging-fiber partially dipped in the fluid [22-24]. The advantage is that the cantilever itself is not damped by the liquid, allowing a precise measurement of the interaction of the fiber with the liquid. Quartz microbalances can also be used to probe liquids at interfaces at MHz frequencies [25] while pump-probe optical spectroscopy methods provide a way to reach GHz frequencies [26].

In this paper, we used the hanging-fiber geometry with three different AFM setups: (i) a commercial AFM setup operated in the frequency-modulation (FM-AFM) mode giving the response of the liquid to an oscillatory excitation in the 50-500 kHz range; (ii) a newly developed AFM based on a microelectromechanical resonator (MEMS-AFM) working at high-frequency around 10-50 MHz [27, 28]; (iii) a homemade high resolution AFM [29] (HR-AFM) used to measure the influence of the liquid interaction on the thermal fluctuations of the cantilever [22, 23]. All methods allow monitoring, independently, the conservative and dissipative contributions of the interaction with the tip. In order to reach precise quantitative information we chose a simple geometry consisting in probes with a micro- or nano- cylindrical fiber tailored at the extremity of an AFM cantilever.

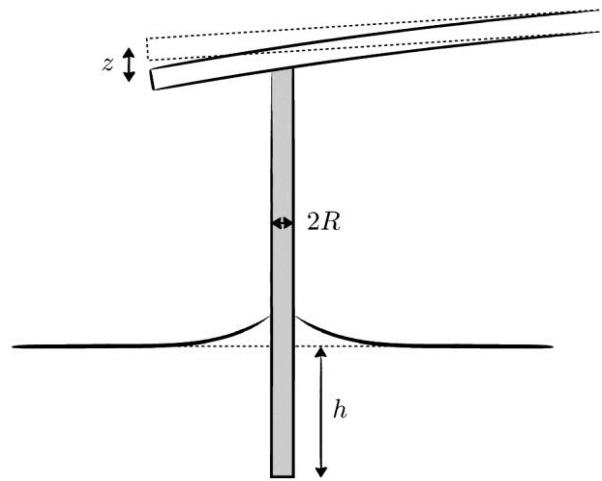

**Figure 1**. Schematic representation of the experiment. The extremity of a fiber with radius $R$ is dipped over a height $h$ with respect to the reference level of the liquid interface while monitoring the $z$ deflection of the cantilever.

In the following, we show that this geometry associated with non-conventional AFM setups allows for quantitative study of the hydrodynamic interaction of a micro or nanosized probe with liquids. A hydrodynamic model is proposed to interpret the whole set of data covering a large range of experimental conditions (probe size, liquid properties) and solicitations, and give a unified picture of the phenomena at stake.

In the first section, we describe the experimental conditions, the different instruments and operating AFM modes, and the raw data that are monitored during the immersion of the probe in the liquid. In the following section, we show how physical quantities can be extracted from the raw data. A theoretical model is then presented and compared to the experimental results.

## EXPERIMENTAL METHODS AND RAW DATA

The tips were dipped in a container drilled in an aluminum or copper sample holder filled with the liquid under study, or in a droplet supported by a silicon substrate, both with diameter ≥ 5 mm and depth ≥ 1 mm. We used a large series of liquids including alkanes, long chains alcohols, glycols, silicon oils and ionic liquids with viscosities ranging from 1 to 1000 mPa.s. The relevant parameters (volumic mass ρ and viscosity η) of the liquids used are listed in Supplemental Material (SM1).

The experiments were performed by using three complementary AFM setups operated in two different modes with micro- and nano-sized probes.

### *Frequency-modulation FM-AFM*

A first series of measurements were performed on a PicoForce AFM (Bruker) operated in the frequency-modulation (FM-AFM) mode using a phase lock loop device (HF2PLL, Zurich Instruments). In this mode, the cantilever is oscillated at one of its resonance frequency (fundamental mode $f_0$ or second mode $f_1$) and the frequency shift $\Delta f$ compared to the oscillation in air is monitored. A PID was used to modulate the excitation signal $A_{ex}$ (in Volt) sent to the piezoelectric element in order to maintain the amplitude of oscillation of the tip constant. The monitoring of the $A_{ex}$ signal gives access to the dissipation of the system. The advantage of the FM-AFM mode compared to the standard amplitude modulation (AM-AFM) mode is that it allows measuring independently the conservative and dissipative parts of the interaction while maintaining the oscillation amplitude constant.

We used two types of AFM tips terminated by a nanocylinder with diameter below 60 nm. These tips are made by focused ion beam milling of a silicon tip (CDP55 by Team Nanotec, Germany) or by growth of an $Ag_2Ga$ nanoneedle at the tip extremity (Nauga Needles, USA). Both types of tips were mounted on cantilevers with a static spring constant of the order of 2 N/m, soft enough to perform static deflection measurements while being adapted for dynamic AFM studies. Measurements were performed both on the fundamental mode with a resonance frequency of the order of *$f_0$* ~ 70 kHz and on the second mode with a resonance frequency *$f_1$ = 6.25 $f_0$* ~ 440 kHz. The associated spring constants, measured using the thermal noise spectrum are *$k_0$* ~ 2 N/m and *$k_1$ = 40 $k_0$* ~ 80 N/m [30].

The results of a typical experiment are plotted in Fig. 2. They are obtained with a silicon tip with a nanocylinder with radius 27.5 nm and length 680 nm (see inset Fig. 2) oscillated at its resonance frequency ($f_0 = 72450\ Hz$ in air) with an amplitude of 7 nm. The tip was dipped in and withdrawn from the liquid bath with a ramp amplitude of 1 μm and a velocity of 2 μm/s. The frequency shift $\Delta f$ and friction coefficient $\beta$ (deduced from $A_{ex}$ as explained below) are reported as a function of the immersion depth $h$ for one series of ionic liquids.

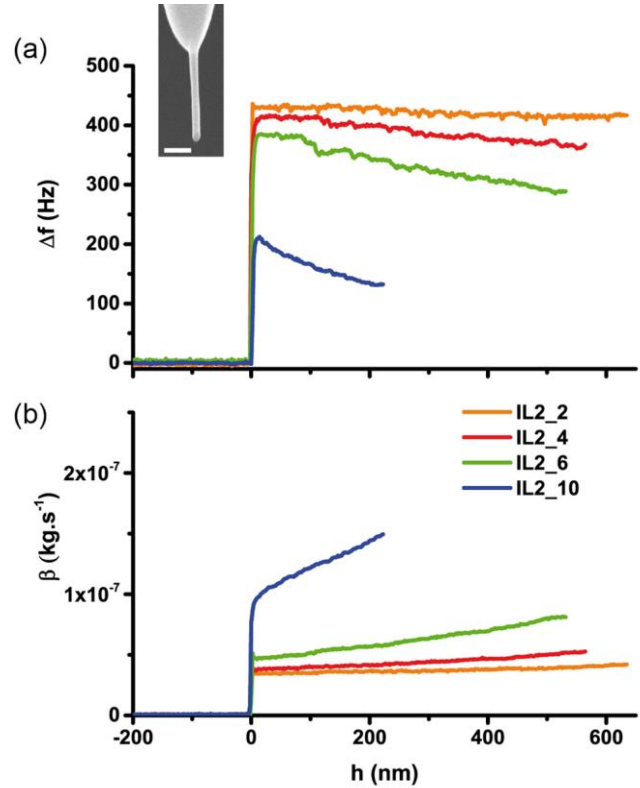

**Figure 2.** (a) Frequency shift and (b) friction coefficient $\beta$ as a function of the immersion depth $h$ for a series of four ionic liquids. Inset: SEM image of the tip used with diameter 55 nm and length 680 nm; scale bar: 200 nm.

Three different signals were monitored during this process:

- The deflection of the cantilever measures the capillary force, which gives information on the wetting properties of the nanocylinder. Since in this paper we do not consider the effects of the meniscus close to the contact line, this curve is not shown. Two plateaus are observed when the tip is dipped in then withdrawn from the liquid bath, corresponding to the advancing and receding contact angles as already discussed in several papers [31-33].

- The frequency shift $\Delta f$ compared to the oscillation in air (Fig. 2a) exhibits a large positive jump when the meniscus is formed. When the tip is dipped further in the liquid we observe a negative slope of the frequency shift which is all the more important that the liquid is viscous. Since the frequency of the cantilever is given by $= 2\pi\sqrt{\frac{k_c}{m_c}}$, the frequency shift may have two origins : a change of the spring constant $\Delta k$ or a change of mass $\Delta m$ according to $\frac{\Delta f}{f} = \frac{1}{2}\frac{\Delta k}{k_c} - \frac{1}{2}\frac{\Delta m}{m_c}$ where $k_c$ and $m_c$ are the spring constant and the effective mass of the cantilever, respectively.

- The normalized excitation $\frac{A_{ex}-A_0}{A_0}$ gives the relative change in excitation $A_{ex}$ required to maintain the amplitude constant compared to the situation in air $A_0$. In order to obtain a quantitative information, the dissipation is characterized by the friction coefficient $\beta = \beta_0 \left(\frac{A_{ex}-A_0}{A_0}\right)$ with $\beta_0 = \frac{k_c}{\omega Q}$ the friction coefficient in air far from the surface [34] where $\omega = 2\pi f$ and $Q$ are the angular frequency and quality factor of the cantilever, respectively. A jump of dissipation in the $\beta(h)$ curve is observed when the meniscus is formed and a positive linear slope is obtained when dipping the nanoneedle further in the liquid. Again, this slope strongly depends on the liquid viscosity.

The same signals can also be recorded on the very same system using the second mode peak of the cantilever of frequency $f_1 = 455400\ Hz$, which allows assessing the influence of excitation rate.

### High-frequency MEMS-AFM

The micro-electromechanical resonators used for the high frequency measurements were designed and fabricated by Walter et al. at IEMN (Lille). Details are reported elsewhere [27, 28]. The resonating element is a ring anchored by 4 points located at vibration nodes (fig. 3a) and is equipped with a sharp pyramidal tip of 5° half angle (fig. 3b). The MEMS device was integrated in a specifically designed home-made AFM microscope (see Supplemental Material SM2 for further details on the setup). The periodic forcing of the resonator and signal acquisition from the microwave detection circuit were performed with a lock-in amplifier including a PLL (HF2LI-PLL from Zurich Instrument). Since the MEMS-AFM was operated in frequency modulation mode, we monitored the same signals, frequency shift $\Delta f$ and excitation amplitude $A_{ex}$, as for the FM-AFM described in the previous section.

The results of a typical experiment are plotted in Fig. 3c-d. They are obtained with a silicon tip with a pyramidal tip oscillated at its resonance frequency ($f_0 = 13,1\ MHz$ in air) with amplitude of 1 nm. The tip was dipped in and withdrawn from the liquid droplet with ramp amplitude of 3 µm and a velocity of 0.3 µm/s. The graphs of frequency shift $\Delta f$ (Fig. 3c) and friction $\beta$ (deduced from $A_{ex}$) (Fig. 3d) are reported as a function of the immersion depth $h$ in Fig. 3c and 3d respectively for a series of ethyleneglycols.

At the meniscus formation, a small negative frequency shift is observed followed by a continuous decrease of the frequency shift upon further dipping of the tip. This latter point is similar to the negative slope observed by FM-AFM. The dissipation signal also follows similar trends as for FM-AFM with a jump in dissipation at the contact with liquid and an increase for positive $h$ values. As discussed in the next section the non linear behavior can be attributed to the pyramidal shape of the tip.

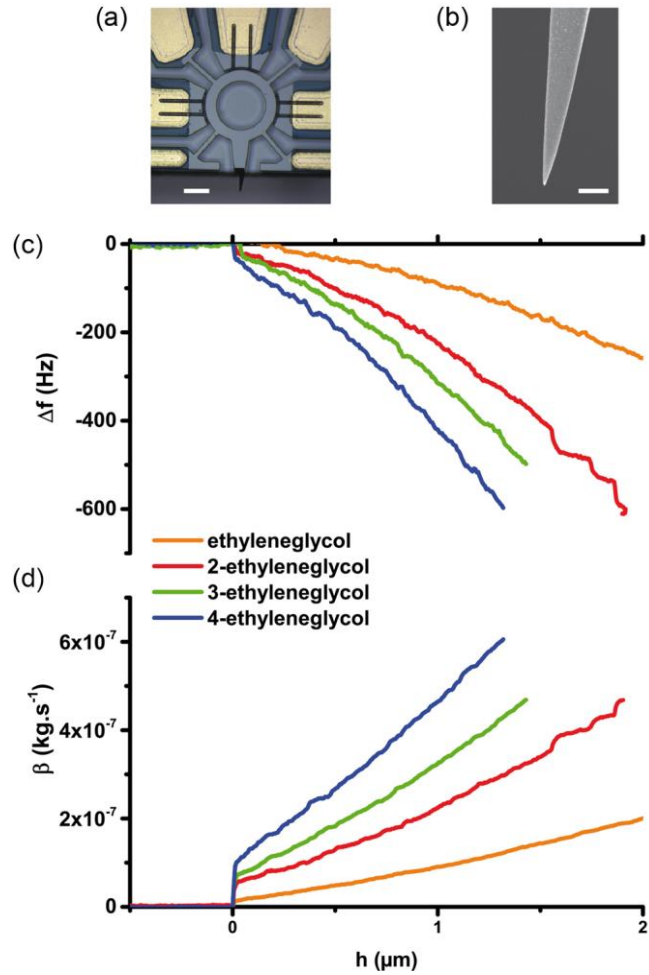

**Figure 3.** (a) Optical micrograph of the resonating element; scale bar: 20 μm. (b) SEM image of the tip extremity; scale bar: 500 nm. (c) Frequency shift $\Delta f$ as a function of the depth $h$ of immersion of the tip for a series of 4 polyethyleneglycols. (d) Same with the friction coefficient $\beta$.

*High-resolution HR-AFM*

The thermal noise spectrum of the deflection $z$ of the cantilever was recorded using a home-built high resolution quadrature phase interferometer, which measures the optical path difference between a laser beam reflecting on the free end of the cantilever above the location of the fiber and a reference beam reflecting on the base of the cantilever [29]. This technique offers a very low detection noise (down to $10^{-14}$ m/√Hz) and is intrinsically calibrated against the laser wavelength. As probes, we used elongated micrometer-sized glass cylinders glued on standard AFM cantilevers (Budget Sensors AIO, levers A and B) and cut between a sharp tweezer and a diamond tool, as described in [22]. These cylinders, typically 1- 10 μm in diameter and 150-250 μm long are glued on a cantilever having a resonant frequency in air $f_0$ ~ 10 kHz (spring constant $k_0$ ~ 0.25 N/m). The Power Spectral Density (PSD) of the thermal fluctuations of the cantilever was measured as a function of the dipping depth $h$, which was varied by 3 or 5 μm steps, allowing the system to relax prior to measurement [22]. For viscosities below or equal to 20 mPa.s, the experimental PSD is fitted around the fundamental resonance peak using a model of simple harmonic oscillator. The data analysis gives access to the frequency shift $\Delta f$ and to the dissipation coefficient $\beta$ associated with the interaction of the tip with the liquid. For viscosities higher than 20 mPa.s, the thermal fluctuations of the cantilever are overdamped. The dissipation coefficient $\beta$ can still be obtained by fitting the experimental PSD by a Lorentzian around the cut-off frequency (see Supplemental Material SM3). However, no information can be obtained on the resonant frequency shift $\Delta f$ in this overdamped regime.

The frequency shift $\Delta f$ with respect to the resonant frequency $f_0$ in air is plotted as a function of the dipping depth $h$ for the same probe in several liquids in Fig. 4a. A negative slope is clearly seen for dipping depths larger than 30 µm, that is to say deep enough for the meniscus to be above the sharp defects created close to the cylinder's end by the cutting procedure (which could be responsible for a local change in the meniscus spring constant). The friction coefficient $\beta(z)$ plotted in Fig. 4b for the same series of liquids shows positive slope. These behaviors are similar to the ones observed for FM-AFM and MEMS-AFM, the general trend being an increase of the slopes with the liquid viscosity.

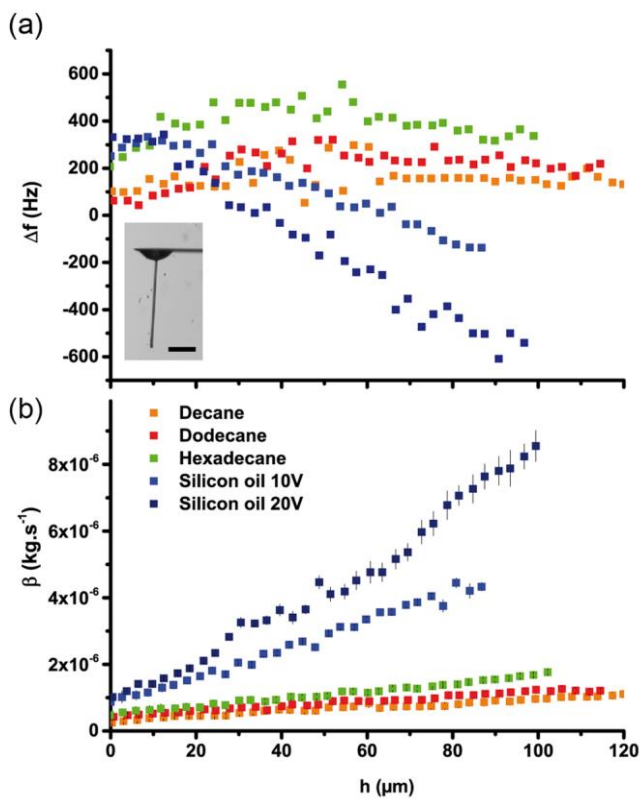

**Figure 4.** (a) Frequency shift and (b) friction coefficient $\beta$ as a function of the immersion depth $h$ for five different liquids. Inset: Optical micrograph of the glass fiber attached to an AFM cantilever; scale bar: 50 µm.

The three AFM techniques therefore give access to similar quantities, namely the frequency shift $\Delta f$ and the friction coefficient $\beta$ obtained in different but complementary conditions: the FM-AFM and MEMS-AFM use nanoprobes with radius in the 20-500 nm with a variable forcing (1-100 nm for FM-AFM and 0.5pm-1nm for MEMS-AFM) whereas thermal noise is a passive method which applies better for micron size probes. FM-AFM and MEMS-AFM monitor in real time both frequency shift and excitation quantities, and therefore the precise evolution of these quantities during the dipping process. On the contrary, thermal noise PSD can be recorded only at given heights $h$ but yields a complete noise spectrum response. Indeed, FM-AFM and MEMS-AFM are well adapted to small variations of quality factor $Q$ on narrow resonance peaks whereas thermal noise is better adapted to strong dissipation leading to small $Q$ values. The association of these techniques gives access to a very large range of conditions in terms of solicitation, probe size and liquid viscosities summarized in table 1. Interestingly, the behavior of the liquid can be probed over four decades of frequency and five decades of amplitude. The radius of the probe can also be changed over two orders of magnitudes whereas the different techniques have rather similar limitations in terms of measurable liquid viscosities.

Note that for FM-AFM and MEMS-AFM modes, we did not observe any influence of the oscillation amplitude in the range 1 pm – 20 nm on the measurements of $\Delta f$ and $\beta$, which indicates that all measurements were performed in the linear regime (see Supplemental Material SM4).

**Table I**. Range of parameters for the three AFM techniques.

| Technique | Forcing | Frequency | Amplitude range | Probe radius $R$ | Liquid viscosity range | Viscous layer thickness $\delta$ |
|---|---|---|---|---|---|---|
| **FM-AFM** | yes | 60 - 500 kHz | 1-100 nm | 20 - 500 nm | 10 - 1000 mPa.s | 5-50 µm |
| **MEMS-AFM** | yes | 13 MHz | 0.5pm – 1 nm | 100 nm - 1µm | 1 – 1000 mPa.s | 0.1-1 µm |
| **HR-AFM** | no | 2kHz – 100 kHz | ~ Å | 500nm - 5µm | 1 - 500 mPa.s | 10-100 µm |

## RESULTS: ADDED MASS AND DISSIPATION COEFFICIENT

In this article, we study the behavior of the liquid around the immersed part of the fiber. With that aim, we concentrate in the following on the slope of the experimental curves (Figs. 2 - 4) obtained for positive $h$. In this region, the measurements can be unambiguously attributed to the effect of the viscous layer since the meniscus around the fiber is independent of the immersed length $h$.

In order to relate the raw data to physical quantities, we write the equation of motion of the tip considered as a point of mass $m_c$ attached to a cantilever with spring constant $k_c$ and submitted in air to a friction coefficient $\beta_{air}$. When dipped in the liquid the system is submitted to the interaction with the fluid which can be divided in two contributions:

The meniscus created around the fiber acts as a spring whose stiffness $k_{men}$ depends on the geometrical shape of the meniscus related to the surface tension and the contact angle [32, 35]. The meniscus also introduces an extra dissipation $\beta_{men}$ which is not discussed here. Given the cylindrical shape of the fiber, these terms do not change during further immersion of the tip in the liquid ($h > 0$ for the nanoneedles and $h > 25\ \mu m$ for the glass microfiber).

Another contribution comes from the liquid around the immersed part of the fiber. The generic expression of the drag force $F_l$ exerted by a liquid on a moving object is the sum of a friction term and an added mass term [36]. The drag force reads:

$$F_l = -\beta^* h.\dot{z} - m^* h.\ddot{z} \qquad (1)$$

where $\beta^*$ and $m^*$ are the friction coefficient and added mass per unit length, respectively.

When these terms are incorporated in the equation of motion of the tip one obtains:

$$(m_c + m^*.h).\ddot{z} + (\beta_{air} + \beta_{men} + \beta^*.h).\dot{z} + (k_c + k_{men}).z = F(t) \qquad (2)$$

where $z$ is the position of the tip with respect to the equilibrium position without oscillation and $h$ is the immersed length of the fiber. $F(t)$ is a harmonic excitation of angular frequency $\omega$ and amplitude $F_0$ for FM-AFM and MEMS-AFM and a delta-correlated random force for thermal noise.

The relative frequency shift is given by $\frac{\Delta f}{f} = \frac{1}{2}\frac{\Delta k}{k} - \frac{1}{2}\frac{\Delta m}{m} = \frac{1}{2}\frac{k_{men}}{k_c} - \frac{1}{2}\frac{m^*.h}{m_c}$. A frequency shift therefore results from a change of stiffness or a change of mass. The positive jump of $\Delta f$ observed in FM-AFM when the meniscus is formed can be attributed to the meniscus spring constant $k_{men}$ [35] whereas the negative slope observed for positive $h$ results from the added mass $m^*.h$ of the viscous layer around the nanofibre. Note that this positive jump is not observed in MEMS-AFM due to the large value of the resonator spring constant compared to the meniscus spring constant ($\frac{k_{men}}{k_c} \sim 10^{-7}$). The added mass per unit length can therefore be calculated using $m^* = -2m_c \left(\frac{\Delta f}{f}\right)^*$ where $\left(\frac{\Delta f}{f}\right)^*$ is the relative frequency shift per unit length $h$.

Accordingly, the shape of the dissipation curves $\beta(h)$ can be interpreted by a jump due to the dissipation in the meniscus $\beta_{men}$ followed by a positive slope resulting from the increase of friction coefficient with the immersed length of the fiber. The friction coefficient per unit length $\beta^*$ is therefore directly measured as the slope of the friction coefficient $\beta$ for positive $h$.

The experimental results presented in the previous section can therefore be used to determine the added mass term $m^*$ and the friction coefficient $\beta^*$. Note that for MEMS-AFM, the tips are not cylindrical but have the shape of a sharp pyramid (half-angle $\alpha = 5°$). The linear increase of the radius with the immersion $h$ leads to an increase of the slopes with $h$. In a rough approximation, the volume of fluid involved with the oscillating motion scales as $\delta.h^2 \tan \alpha$ where $\delta$ is the thickness of the liquid layer dragged by the probe. In the following, we therefore measured the slope of the $m^*(h)$ and $\beta^*(h)$ curves, just after the meniscus formation ($h = 0$). The corresponding wetted radius $R$ were deduced from the meniscus height which is obtained from the hysteresis between advancing and retracting curves [33].

The added mass $m^*$ and friction coefficient $\beta^*$ were deduced from the FM-AFM (for both fundamental and second modes), MEMS-AFM and HR-AFM experiments on a large number of liquids. Examples of results are reported on Fig. 5. For clarity, we only reported several series of liquids. The whole set of measurements will be presented in section V.

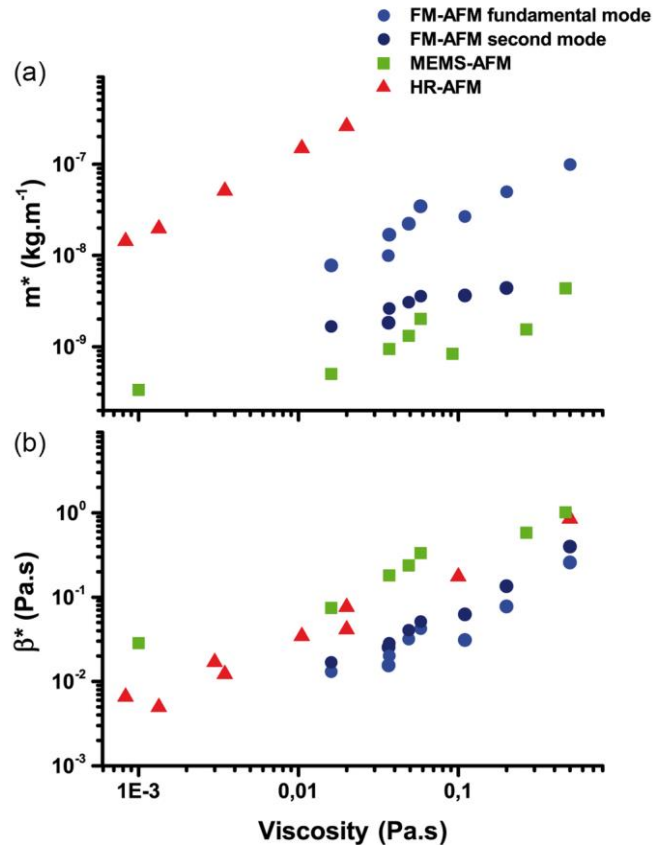

**Figure 5.** (a) Added mass $m^*$ and (b) friction coefficient $\beta^*$ extracted from the three types of experiments, reported as a function of the liquid viscosity.

A general trend observed on Figure 5 is that, for a given experiment, *i.e.* for fixed probe radius and excitation frequency, both $m^*$ and $\beta^*$ quantities increase with the liquid viscosity. The results also demonstrate that the values strongly depend on the type of AFM experiment. Since both probe radius $R$ and excitation angular frequency $\omega$ vary from experiment to experiment, their respective influences on the measurements are not straightforward. However, FM-AFM can give an indication on the influence of frequency by comparing the response of the very same system (tip + liquid) to solicitations at the fundamental and second modes which differs by a factor 6.25 in frequency. It appears that frequency has a small effect on friction coefficient $\beta^*$ whereas the added mass term $m^*$ decreases by a factor of the order of 7 when increasing the frequency. This influence of excitation frequency may also explain the difference of 3 orders of magnitude between $m^*$ values obtained by MEMS-AFM and HR-AFM. On the contrary, these two techniques lead to similar $\beta^*$ values, significantly larger than the ones deduced from FM-AFM experiment.

In order to interpret these results and give a unified vision of the whole data by disentangling the influence of the different parameters, we developed a model which is presented in the next section.

# THEORETICAL MODEL

In this section, we establish the theoretical framework in which the hydrodynamic problem at stake can be understood. The hydrodynamics of rod-like particles has attracted a lot of attention due to its implication in many important issues such as the rheological properties of solutions of such objects (polymers, nanotubes), or in size measurements from light scattering experiments [2, 4]. Approximate solutions have been proposed for ellipsoids and cylinders of given aspect ratio totally immersed in an infinite liquid bath [4, 37]. Since in our case, the effects of the meniscus and of the immersed end may be considered as constant during the dipping process, we considered the case of an infinite cylinder oscillating longitudinally in an infinite liquid bath. The latter assumption is justified since in the experiments the diameter of the liquid vessel or droplet (6 mm for FM-AFM and MEMS-AFM, 14 mm for HR-AFM) is much larger than the penetration depth of the shear waves in the liquids (of order of 5-50 µm). With that aim we revisited the model established by Batchelor for a cylinder moving steadily in a viscous liquid [38].

In order to compute the longitudinal drag force, we considered the flow induced by an infinite cylindrical rod oscillating along its axis in a purely viscous Newtonian liquid. The rod's radius is denoted by $R$ (see Fig. 1) and its speed, assumed to be harmonic (with no lack of generality thanks to the linearity of the equations), reads $v_0\,e^{-i\omega t}$. The velocity field $v(r,t)$ in the liquid is obtained by solving the cylindrical Stokes equation [36] $\varrho\,\partial_t v = \eta\left(\partial_r^2 v + \frac{\partial_r v}{r}\right)$, together with a no-slip boundary condition at the rod's surface $v(R,t) = v_0\,e^{-i\omega t}$, and a vanishing-speed boundary condition at infinity $\lim_{r\to\infty} v(r,t) = 0$. The solution reads for all $r \geq R$:

$$v(r,t) = v_0\,e^{-i\omega t}\,f\left(\frac{r}{\delta}, \frac{R}{\delta}\right) \qquad (3)$$

where $\delta = \sqrt{\frac{2\eta}{\rho\omega}}$ is the skin thickness of the well-known 2D case of a viscous flow induced by an oscillating plate in a liquid [36], and the function of two variables $f$ reads:

$$f(u_1, u_2) = \frac{\mathcal{K}_0[(1-i)\,u_1]}{\mathcal{K}_0[(1-i)\,u_2]} \qquad (4)$$

where $\mathcal{K}_n$ denotes the modified Bessel function of the second kind of order *n* [39]. The real velocity field thus oscillates in an envelope $v_e$ (see Fig. 6), whose derivation and expression is given in supplementary information SI5. When the radius $R$ is of order, or smaller than $\delta$, the velocity field becomes more confined than in the 2D solution leading to an effective skin smaller than $\delta$. This effective skin depth $\delta_{\text{eff}}$ will be evaluated quantitatively in the following. When $R \gg \delta$, the velocity profiles match that of the aforementioned 2D case. In the latter limit, the influence of the cylindrical geometry can thus be safely neglected.

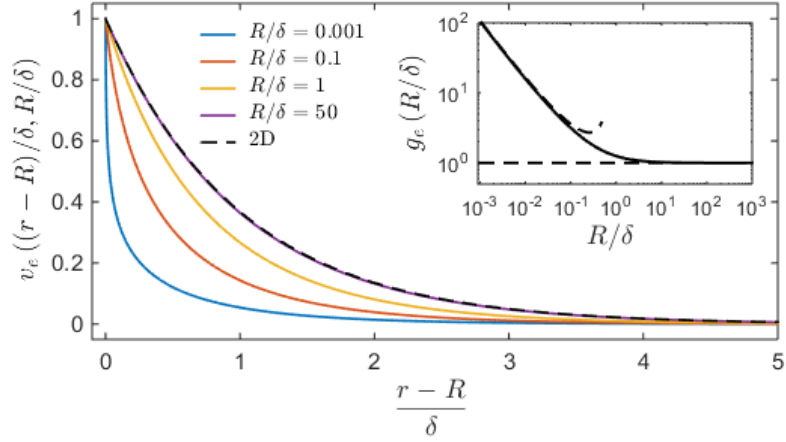

**Figure 6.** Plot of the velocity envelope $v_e[(r-R)/\delta, R/\delta]$ as a function of $(r-R)/\delta$ for different values of $R/\delta$. The black dashed curve corresponds to the well-known 2D viscous flow induced by an oscillating plane [36]. Inset: Plot of $g_e\left(\frac{R}{\delta}\right)$, envelope of the normalized shear stress $g\left(\frac{R}{\delta}\right)$, given by Eq. (6). Dashed lines correspond to the asymptotic solutions for $R \gg \delta$ and $R \ll \delta$ (Eq. 7).

The frictional force on the rod per unit area is given by the shear stress at the rod's surface:

$$\sigma = -\sigma_{rz}|_{r=R} = -\eta\ \partial_r v|_{r=R} = \frac{\eta}{\delta}(1-i)\ g\left(\frac{R}{\delta}\right) e^{-i\omega t} \qquad (5)$$

where the function g reads [39]:

$$g(u) = \frac{\mathcal{K}_1[(1-i)\ u]}{\mathcal{K}_0[(1-i)\ u]} = g_1(u) + i \times g_2(u) \qquad (6)$$

with $g_1$ and $g_2$ being respectively the real and imaginary parts of the function g.

In Fig. 6, one can notice that as the radius R becomes small compared to the 2D skin thickness δ, the velocity gradient at the rod's surface increases dramatically. The shear stress oscillates with an amplitude $\sigma_e\left(\frac{R}{\delta}\right) = \frac{\eta}{\delta}\ g_e\left(\frac{R}{\delta}\right)$. The stress enhancement factor $g_e\left(\frac{R}{\delta}\right)$ plotted in the inset of Fig. 6 (expression given in Supplemental Material SM5) diverges for small R/δ as:

$$R/\delta \to 0\ g_e\left(\frac{R}{\delta}\right) \sim \frac{-1}{\sqrt{2}\times\frac{R}{\delta}\times \ln\left(\frac{R}{\delta}\right)} \qquad (7)$$

In the case of the FM-AFM experiments for which the ratio $\frac{R}{\delta}$ is of order $10^{-3}$, an enhancement of the stress by a factor 100 is found with respect to a planar situation. This leads to large shear rates of the order of $10^5$ s$^{-1}$ to $10^6$ s$^{-1}$ in standard conditions.

Finally, the frictional force per unit length along the rod axis reads $F^* = 2\pi R\, \sigma$. Using Eq. (5), the force can be written in a temporal form equivalent to Eq. (1) as:

$$F^* = -\beta^*\ v|_{r=R} - m^*\ \partial_t v|_{r=R} \qquad (8)$$

where $\beta^*$ and $m^*$ are respectively a friction coefficient and a mass term, both per unit length. They read:

$$\beta^* = 2\pi\eta \times \frac{R}{\delta}(g_1 + g_2) = 2\pi\eta \times C_\beta \tag{9}$$

and

$$m^* = 2\pi\frac{\eta}{\omega} \times \frac{R}{\delta}(g_1 - g_2) = 2\pi\frac{\eta}{\omega} \times C_m \tag{10}$$

For high values of $R/\delta$ friction coefficient and added mass are related through a simple expression which is, as one might expect, that of the 2D situation:

$$R/\delta \to +\infty \qquad \beta^* \sim m^* \times \omega \tag{11}$$

The coefficients $C_\beta$ and $C_m$ are represented in Figure 7 as function of $R/\delta$. For large $R/\delta$ values, $C_\beta$ and $C_m$ meet the same asymptotic behaviour $C_\beta \sim C_m \sim R/\delta$. The friction coefficient in this regime reads: $\beta^* \sim m^* \cdot \omega \sim \pi R\sqrt{2\eta\rho\omega}$. For small $R/\delta$, the asymptotic behavior of $C_\beta$ and $C_m$ reads:

$$R/\delta \to 0 \quad C_\beta \sim \frac{-1}{\ln(\frac{R}{\delta})} \quad \text{and} \quad C_m \sim \frac{\pi/4}{[\ln(\frac{R}{\delta})]^2} \tag{12}$$

In this regime, the coefficient $C_\beta$ is weakly dependent on the probe size, liquid viscosity $\eta$ and excitation frequency $\omega$ leading $\beta^* \sim -2\pi\eta\ln(\frac{R}{\delta})$. The mass term reads $m^* \sim -2\pi\frac{\eta}{\omega}\left[\ln(\frac{R}{\delta})\right]^2$.

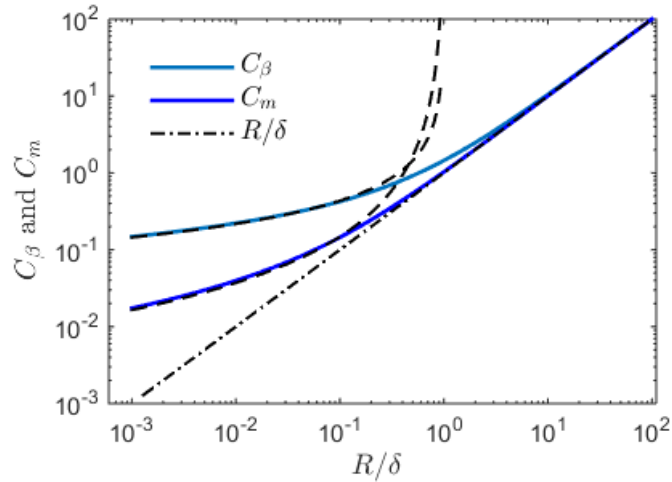

**Figure 7.** Plot of $C_\beta$ and $C_m$ (see Eq. (9) and (10)) as a function of $R/\delta$ (solid curves). The dashed black curves display the asymptotic regime as given by Eq. (12) and the dotted and dashed black curve displays the 2D regime.

We may now use the previous results to evaluate the aforementioned effective skin depth $\delta_{eff}$, by analogy with the 2D situation. In this latter case indeed, the added mass is half the mass of the fluid localized in the viscous layer of depth $\delta$. One shall take therefore $\delta_{eff}$ such as the mass per unit length of an annular section of liquid with inner radius $R$ and outer radius $R + \delta_{eff}$ is equal to $2m^*$. Identifying with Eq. (10), one has:

$$\frac{\delta_{eff}}{\delta} = -\frac{R}{\delta} + \sqrt{\left(\frac{R}{\delta}\right)^2 + 2 \times C_m} \qquad (13)$$

Figure 8 represents the ratio $\delta_{eff}/\delta$ as a function of $R/\delta$. For large values of $R/\delta$, one recovers as expected the 2D regime for which $\delta_{eff} = \delta$ (dashed red curve). When $R$ is of order of, or smaller than $\delta$, the effective skin length $\delta_{eff}$ decreases and reaches the small radius asymptotic regime (dashed black curve), for which $R$ can be neglected in the calculation of the annular section area, leading to:

$$R/\delta \to 0 \quad \frac{\delta_{eff}}{\delta} \sim \sqrt{2\,C_m} \sim \frac{-\sqrt{\pi/2}}{\ln(\frac{R}{\delta})} \qquad (14)$$

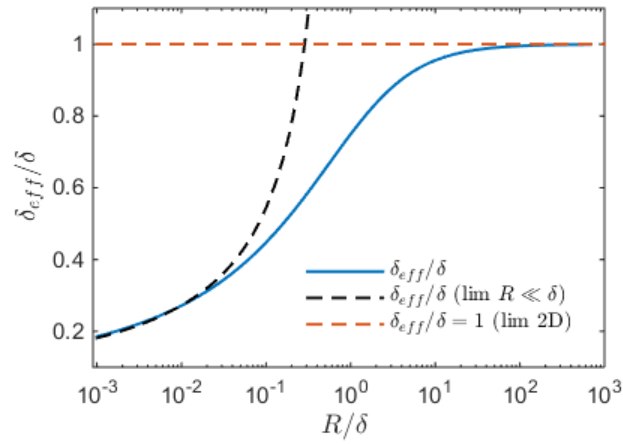

**Figure 8.** Plot of $\delta_{eff}/\delta$ as given by Eq. (13) as a function of $R/\delta$ (solid blue curve). The dashed red line displays the 2D regime for which $\delta_{eff} = \delta$, and the dashed black curve displays the small radius asymptotic regime as given by Eq. (14).

Equation (13) gives a quantitative description of the extension of the velocity field around a nanocylinder. For probes with cylinder radius much smaller than the thickness of the 2D viscous layer $\delta$, the extension of the viscous layer is significantly reduced. One observe that $\frac{\delta_{eff}}{\delta}$ decreases dramatically when $R/\delta$ decreases from $10$ to $10^{-2}$ (with a drop of approximately 80%). Further decrease of the probe size has small (logarithmic) effect.

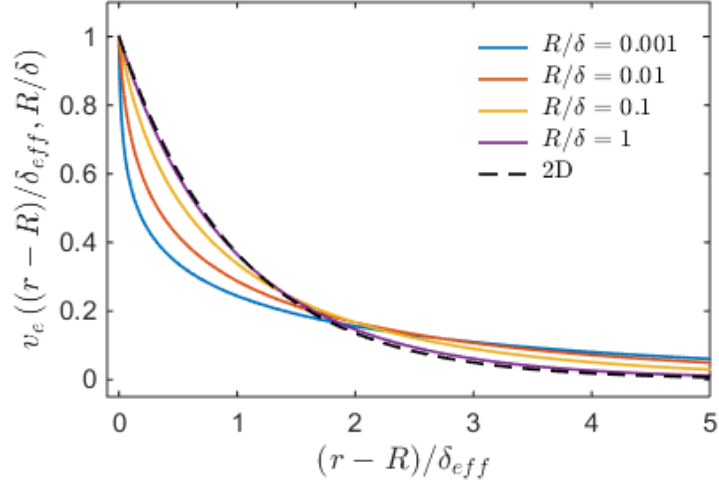

**Figure 9.** Plot of the velocity envelope $v_e[(r-R)/\delta_{eff}, R/\delta]$ as a function of the rescale variable $(r-R)/\delta_{eff}$ for different values of $R/\delta$. The black dashed curve corresponds to the 2D viscous flow induced by an oscillating plane.

Contrary to the friction coefficient $\beta^*$ which is only sensitive to the shear stress at the solid-liquid boundary, the added mass takes into account the whole velocity profile. In particular, in the limit $R/\delta \to 0$ the friction coefficient decreases less strongly than the added mass. This can be understood using a simple handwaving argument. Figure 9 displays the velocity profiles as a function of the rescaled variable $(r-R)/\delta_{eff}$. In this representation, the rescaled added mass does not vary with $R/\delta$ while the rescaled shear stress at the solid-liquid boundary still significantly increases in amplitude as $R/\delta$ is decreased. This interesting feature which explains the difference between $C_\beta$ and $C_m$ for $R/\delta \ll 1$ is inherent to the cylindrical geometry of the system. When $\frac{R}{\delta} > 1$ the normalized curves of Fig. 8 superpose and one regains a quasi-2D geometry and recovers the well-known relation $\beta = m \times \omega$.

## DISCUSSION

The model described above provides a comprehensive description of the behavior of the viscous layer as a function of the experimental conditions. It was shown that different regimes may occur as a function of the relative values of the probe size $R$ and the relevant lengthscale $\delta$ of the problem which depends on the liquid properties $\eta$ and $\rho$ and the excitation angular frequency $\omega$ through $\delta = \sqrt{\frac{2\eta}{\rho\omega}}$. Typical values of $R$ and $\delta$ are reported in Table 1. It shows that the small probe diameters used in FM-AFM, significantly smaller than the thickness $\delta$, provide a way to assess the regime of small $\frac{R}{\delta}$ ($\frac{R}{\delta} \sim 6.10^{-4} - 2.10^{-2}$) where geometrical aspects are important. In MEMS-AFM, the high frequencies lead to small sub-micrometric $\delta$ values which may be comparable with the probe size. HR-AFM uses large probes whose radius is also of the same order as $\delta$. These last cases therefore allows to access an intermediate regime approaching the 2D case ($\frac{R}{\delta} \sim 1.10^{-2} - 1$).

The fact that MEMS-AFM and HR-AFM give access to the same range of $\frac{R}{\delta}$ values explains that these techniques lead to similar values of the friction coefficient $\beta^*$. Since $C_\beta$ is an increasing function of $\frac{R}{\delta}$, these values are also expected to be larger than the ones from FM-AFM as observed on Fig. 5. The same trends are expected for $m^*\omega$, which is consistent with the strong influence of the excitation frequency on the added mass term discussed in section III.

In order to assess more quantitatively the model, the whole set of experimental results was compared to the model using the master curves defined by Eq. 9 and 10 and represented in Fig. 10. In this representation, we plotted for each experimental data point the value of the coefficient $C_\beta$ and $C_m$ defined as $C_\beta = \frac{\beta^*}{2\pi\eta}$ and $C_m = \frac{m^*.\omega}{2\pi\eta}$ as a function of the ratio $\frac{R}{\delta}$. Large values of $\frac{R}{\delta}$ corresponds to the 2D situation whereas the cylindrical geometry needs to be taken into account for $\frac{R}{\delta} \ll 1$, as sketched in the inset of Fig. 10. Interestingly, the large range of probe size, solicitation conditions and liquid viscosities provided by the combination of all AFM techniques allows to vary the $\frac{R}{\delta}$ parameter over more than three orders of magnitude, which corresponds to 6 orders of magnitude in Reynolds number.

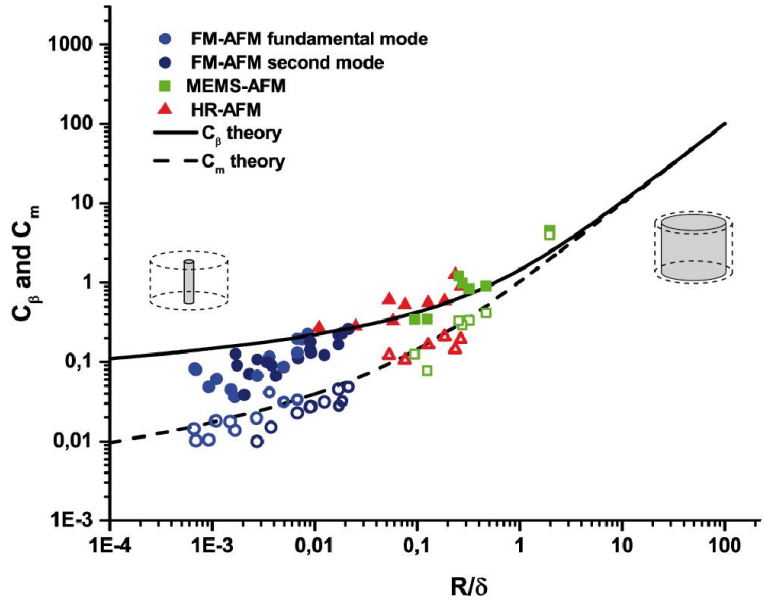

**Figure 10.** Master curve of the $C$ coefficient determined as $C_\beta = \frac{\beta^*}{2\pi\eta}$ (full symbols) or $C_m = \frac{m^*.\omega}{2\pi\eta}$ (empty symbols) as a function of $R/\delta$ for the whole set of experiments with FM-AFM (fundamental mode in light blue circles, second mode in dark blue circles), MEMS-AFM (green squares) and HR-AFM (green triangles). The full line corresponds to the theoretical value values of $C_\beta$ expressed in Eq. 9, the dashed line to the one of $C_m$ expressed in Eq. 10.

The whole set of experimental data points are reported in Fig. 10 where filled symbols and solid line corresponds to experimental and theoretical values of the dissipation $C_\beta$ parameter whereas open symbol and dashed line represents the mass $C_m$ parameter. The colors are associated with the measurement techniques namely, blue (dark blue) for fundamental (second) mode in FM-AFM, red for HR-AFM and green for MEMS-AFM. We observe that each series of points ($C_\beta$ and $C_m$) gather well on a single curve demonstrating the pertinence of the $\frac{R}{\delta}$ parameter to describe the liquid behavior. All data also show a good agreement with the model. More precisely, the mass term is very well reproduced by the model. This is also the case for the dissipation data except for the points corresponding to $\frac{R}{\delta} < 5.10^{-3}$ which were obtained by FM-AFM on the most viscous liquids. In this case, the experimental data are overestimated by about 30-50% by the model. The origin of this discrepancy remains unclear. No particular issue can be anticipated from the measurement side. Some assumptions such as the no-slip boundary condition could potentially be invalidated at the nanoscale. The calculations can be revised with a Navier-like slip boundary condition which has the effect of reducing the shear stress and thus reducing the friction coefficient. Interestingly, the points which corresponds to low $\frac{R}{\delta}$ values were obtained with ionic liquids which exhibit a strong structuration at solid surface as demonstrated by surface force apparatus [40]. The strong stress enhancement due to geometrical aspects as discussed in the previous section may induce a slip in this particular case.

Since no adjustable parameter is present in the theory, this study demonstrates that the experimental protocol together with the methods used to interpret the data and obtain physical parameters such as added mass and friction coefficient are robust. It establishes that FM-AFM, MEMS-AFM and HR-AFM are techniques which can be used with confidence for quantitative investigation of dissipation processes at micro- and nano-scales.

According to the description in terms of effective thickness of the viscous layer discussed above [Eq. (13) and Fig. 8], the experiments performed in FM-AFM mode correspond to values of the effective thickness of the order of $\delta_{\text{eff}} \sim 0.2\,\delta$ whereas for MEMS-AFM and HR_AFM the dynamic confinement is moderate with $\delta_{\text{eff}} \sim 0.5 - 0.8\,\delta$. This quantifies the intuitive fact that, for a small tip radius, the velocity field extends less in the liquid than for the 2D situation of an oscillating plane. Even if the reduction of the probe dimension limits the extension of the velocity field compared to the 2D situation, this reduction remains rather small even for tip radius which are 1000 smaller than the thickness of the viscous layer. The fact that the velocity field extends over several microns may have important implications for the collective motion of passive or active nanoparticles in a liquid environment. The effective thickness of the viscous layer are therefore of the order of $\delta_{\text{eff}} \sim 1 - 10\,\mu m$, $\delta_{\text{eff}} \sim 0,1 - 1\,\mu m$ and $\delta_{\text{eff}} \sim 5 - 50\,\mu m$ for FM-AFM, MEMS-AFM and HR-AFM respectively.

# CONCLUSION

In this article, we have shown that AFM techniques, combined with a model tip geometry are powerful tools for the quantitative study of hydrodynamics in a viscous layer around a micro- or nano- cylindrical probe. We implemented three different methods and defined protocols allowing the independent measurement of the added mass and friction terms over a large range of probe size, fluid viscosity and excitation frequency. A model was developed to account for the experimental observations. It shows that the relevant parameter is the ratio $R/\delta$ of the probe size $R$ to the 2D viscous layer thickness $\delta$. All experimental data can be gathered on two master curves, one for added mass and one for friction coefficient, using $R/\delta$ ratio as the single control parameter spanning over three orders of magnitude. They are quantitatively reproduced by the theoretical model without any adjustable parameter, showing the potential of AFM to measure dissipation processes in liquids down to the nanometer scale.

The theoretical model allows quantifying the extension of the velocity field around the nanoprobe and shows it is significantly lower than the thickness of a 2D viscous layer $\delta$. This confinement is associated with a strong enhancement of the surface stress as the probe size is decreased. The scaling laws provided by the model are of interest for the development of nanorheology, in particular for complex fluids which may exhibit nano- to micro-scale characteristic lengths. In this context, the MEMS-AFM experiment allows probing small liquid volumes with characteristic length in the 100 nm and is also relevant in the recent field of high frequency nanofluidics [18, 26, 41].

The quantitative measurement of dissipation processes in liquid around a nanometer scale probe provided by this study may also open the way for systematic investigation of dissipation in intrinsically small liquid volumes such as nanomenisci or at contact lines, issues which are not fully understood despite their great importance in wetting science [42].

# ASSOCIATED CONTENT

*Acknowledgements*

We thank Elie Raphaël and Sergio Ciliberto for fruitful discussions, Philippe Demont for his help in the measurements of ionic liquids viscosity and Christopher Madec for contribution to the HR-AFM experiment. This study has been partially supported through the ANR by the NANOFLUIDYN project (grant n° ANR-13-BS10-0009) and Laboratory of Excellence NEXT (grant n° ANR-10-LABX-0037) in the framework of the "Programme des Investissements d'Avenir". Financial support of the ERC project OUTEFLUCOP is also acknowledged.

*Author information*

Corresponding Author : Thierry Ondarçuhu

*E-mail: ondar@cemes.fr

# Supplemental Material

*Content*

SM1: Chemical structure and characteristic properties of the liquids used

SM2: MEMS-AFM

SM3: HR-AFM on viscous liquids

SM4: Effect of oscillation amplitude on FM-AFM and MEMS-AFM

SM5: Velocity profile and shear stress

# SM1: Chemical structure and characteristic properties of the liquids used.

| Name | Density (kg.m$^{-3}$) | **Viscosity (Pa.s) at 25°C** |
|---|---|---|
| **Decane** | 730 at 20°C | **0.00085** |
| **Dodecane** | 748 at 20°C | **0.00136** |
| **Hexadecane** | 773 at 20°C | **0.00305** |
| **Silicon oil 10v** | 930 | **0.01** |
| **Silicon oil 20v** | 950 | **0.02** |
| **Silicon oil 100v** | 965 | **0.100** |
| Mineral oil M500 | 885 at 20°C | **0.588** (20°C) |
| **Ethyleneglycol** | 1.113 | **0,016** |
| **Diethyleneglycol** | 1.118 | **0,037** |
| **Triethyleneglycol** | 1.124 | **0,049** |
| **Tetraethylene glycol** | 1.125 | **0,058** |
| Octanol | 0.827 | **0,0076** |
| Nonanol | 0.827 | **0,01** |

| | | | | |
|---|---|---|---|---|
| **Decanol** | | | 0.829 | **0,012** |
| **Undecanol** | | | 0.83 | **0,014** |
| **IL1-2** | 1-Ethyl-3-MethylImidazolium | Ethylsulfate | 1239 at 20°C | **0.0925** |
| **IL1-6** | 1-Ethyl-3-MethylImidazolium | Hexylsulfate | 1100 at 25°C | **0.266** |
| **IL1-8** | 1-Ethyl-3-MethylImidazolium | Octylsulfate | 1095 at 20°C | **0.468** |
| **IL2-2** | 1-Ethyl-3-MethylImidazolium | Tetrafluoroborate | 1294 at 25°C | **0.0365** |
| **IL2-4** | 1-Butyl-3-MethylImidazolium | Tetrafluoroborate | 1200 at 20°C | **0.110** |
| **IL2-6** | 1-Hexyl-3-MethylImidazolium | Tetrafluoroborate | 1150 at 20°C | **0.200** |
| **IL2-10** | 1-Decyl-3-MethylImidazolium | Tetrafluoroborate | 1070 at 20°C | **0.500** |

## SM2: MEMS-AFM

The micro-electromechanical resonators used for the high frequency measurements were designed and fabricated by Walter et al. at IEMN (Lille). The resonating element is a ring anchored by 4 points located at vibration nodes and is equipped with a pyramidal tip. The MEMS device was integrated in a specifically designed home-made AFM microscope, which is shown in Fig. Si1. The silicon chip supporting the microfabricated probe was glued with silver epoxy on the tip of a V-shaped mini-PCB (inset Fig. SI1), and connected by ball bonding to gold HF-compatible transmission lines through 25 um gold wires. The mini-PCB was equipped with two HF connectors, one for the resonator excitation and another for the microwave detection. The probe holder was screwed to a L-shaped aluminum block that was fixed to a XYZ precision stage (Newport 562-XYZ). The Z-axis displacement dedicated to the surface approach toward the tip (Fig. SI2) was performed with a stepper motor actuator (Newport TRA12PPD). The periodic forcing of the resonator and signal acquisition from the microwave detection circuit were performed with a lock-in amplifier including a PLL (HF2LI-PLL from Zurich Instrument). The AFM scanner was a 5 $\mu$m × 5 $\mu$m × 5 $\mu$m PicocubeTM piezo scanner (P-363, PI) driven by the E-536 controller (PI). A RC4 Nanonis AFM controller was used for driving the instrument, with a Labview-based modular software that allows automatic surface approach, feedback loop control and Z-spectroscopy.

The characteristics of the MEMS were k=2.10$^5$ N/m, Q=500, f$_0$=13.1 MHz.

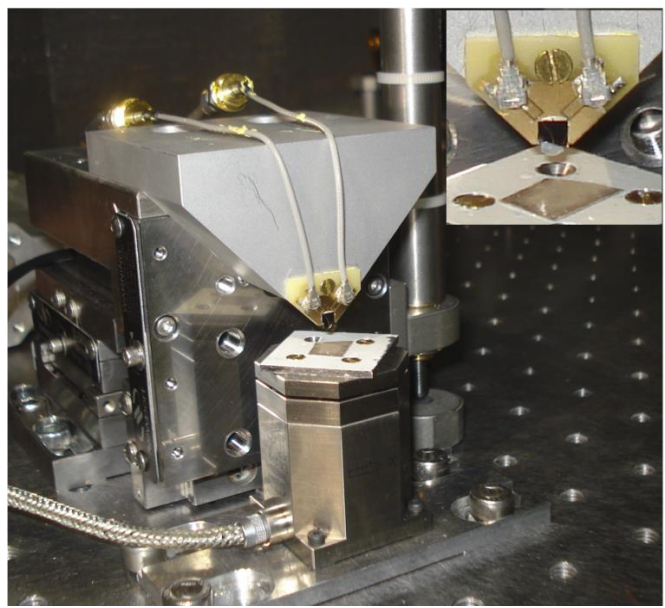

**Figure SM1.** Photo of the MEMS-AFM. Inset: Mini-PCB supporting the MEMS chip.

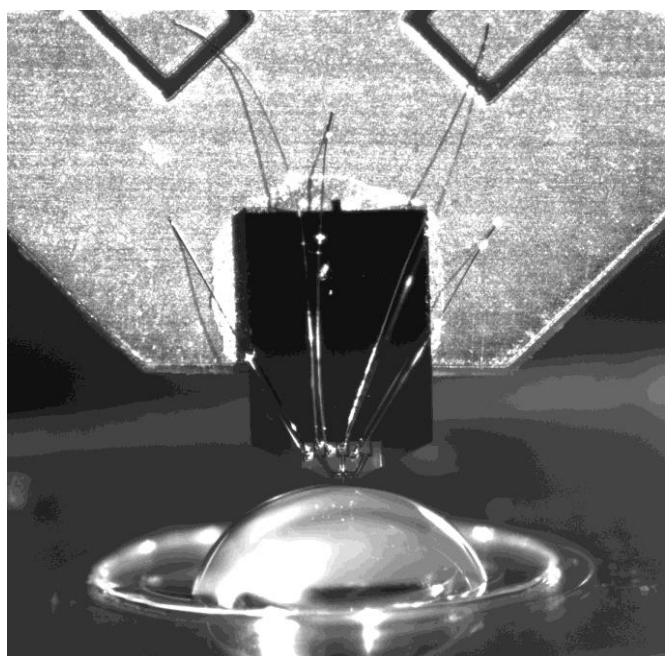

**Figure SM2.** MEMS chip on top of an ethylene glycol droplet. The step motor allows approaching the drop before performing Z-spectroscopy with the Picocube scanner Z-axis.

# SM3: HR-AFM on viscous liquids

In order to analyze the HR-AFM data, we fit the experimental power spectrum density (PSD) of the cantilever thermal vibrations using a simple harmonic oscillator model with viscous damping. The PSD of the thermal vibrations of a simple harmonic oscillator with a spring constant $k$, mass $m$ and viscous friction $\beta$ at temperature $T$ reads:

$$S(f) = \frac{4 k_\text{B} T}{k 2\pi f_c \left[ \left( 1 - \left(\frac{f}{f_0}\right)^2 \right)^2 + \left(\frac{f}{f_c}\right)^2 \right]} \tag{SI1}$$

where $k_\text{B}$ is the Boltzmann constant, $f_0 = \frac{1}{2\pi}\sqrt{\frac{k}{m}}$ is the resonance frequency and $f_c = \frac{1}{2\pi}\frac{k}{\beta}$ is the cut-off frequency. When $f_c \ll f_0$, that's to say $\beta \gg \sqrt{km}$, the vibrations are overdamped and equation (SI1) reduce to a Lorentzian:

$$S_\text{Lor}(f) = \frac{4 k_\text{B} T}{k 2\pi f_c \left[ 1 + \left(\frac{f}{f_c}\right)^2 \right]} \tag{SI2}$$

In this overdamped regime, no information can be recovered on the resonant frequency $f_0$ nor on the mass $m$ from the PSD.

For low viscosity liquids, for which the cantilever stays in a resonant regime over the whole range of dipping depths, the experimental PSD are fitted using equation (SI1) around the fundamental resonance frequency, avoiding the low-frequency region where the coupling with the hanging fiber thermal vibrations affects the measurements (Devailly et al., 2014). For data sets where the cantilever goes from a resonant to an overdamped regime when increasing the dipping depth (see figure SI3, left panel), we keep fitting the whole set of data using equation (SI1), however the $f_0$ parameter becomes meaningless for overdamped curves. We have checked on a few overdamped test curves that the $k$ and $f_c$ parameters obtained by fitting the same data with equation (SI1) and (SI2) were compatible. Finally, for high viscosity liquids where the cantilever is overdamped for all dipping depth, the data are fitted using equation (SI2) around the cut-off frequency. The error bars on the fitting parameters are evaluated by modifying the fitting range.

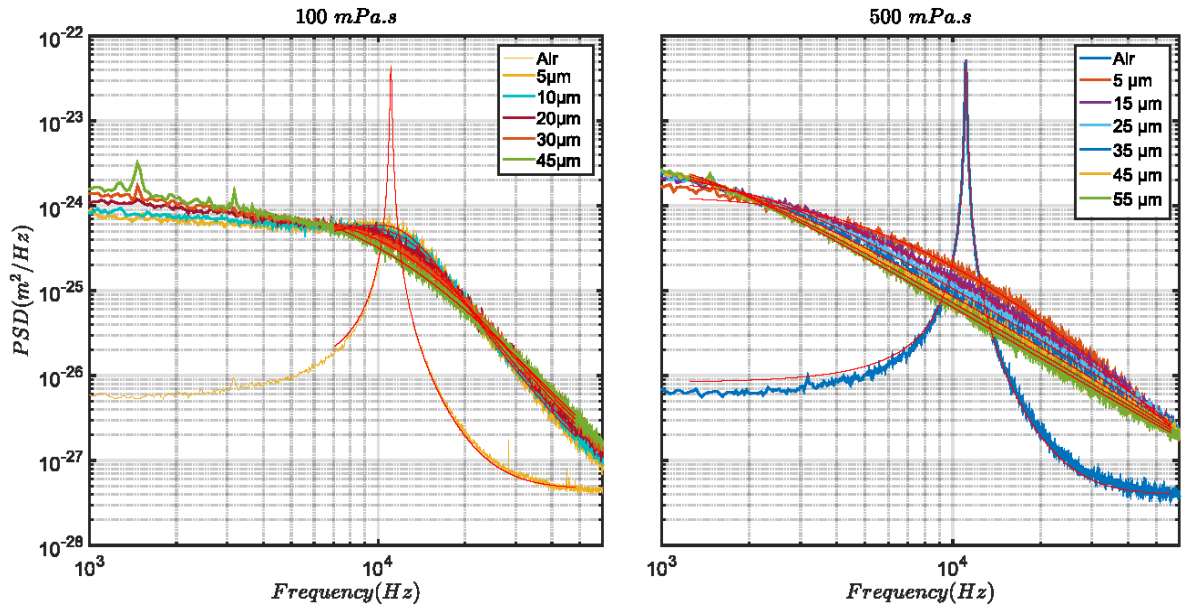

**Figure SM3.** PSD of the thermal fluctuations of a cantilever with a 3.85 μm in diameter fiber glued at its extremity, in air and dipped at different immersion depths in silicone oil 100v (left) and mineral oil M500 (right). Fits using equation (SI1) (right panel) and (SI2) (left panel) are displayed as thin red lines.

## SM4: Effect of oscillation amplitude on FM-AFM and MEMS-AFM

The influence of oscillation amplitude was assessed in the FM-AFM mode. As shown on Fig. SI4, $\Delta f(h)$ and $\beta(h)$ curves do not depend on the oscillation amplitude from 1 nm to 21 nm. Deviation which appears for larger amplitudes (not shown) can be attributed to dissipation at the contact line which does not remain anymore pinned for large solicitations.

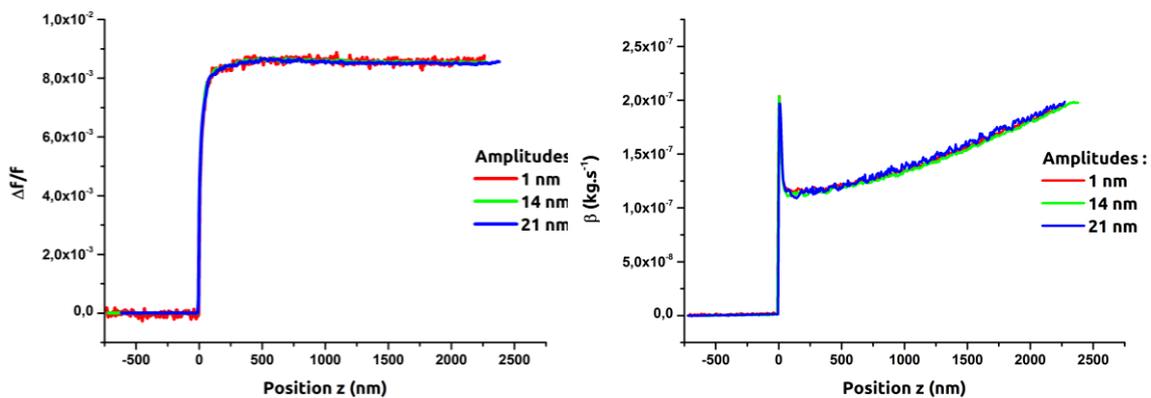

**Figure SM4.** Relative frequency shift (left) and friction coefficient (right) measured by FM-AFM as a function of the oscillation amplitude.

The influence of oscillation amplitude was also investigated for the MEMS-AFM. In Figure S5 are reported the $\beta(h)$ curves for amplitudes of 0,5 nm and 0,5 pm. The latter value is the lower limit available with the MEMS-AFM which explains the large noise level.

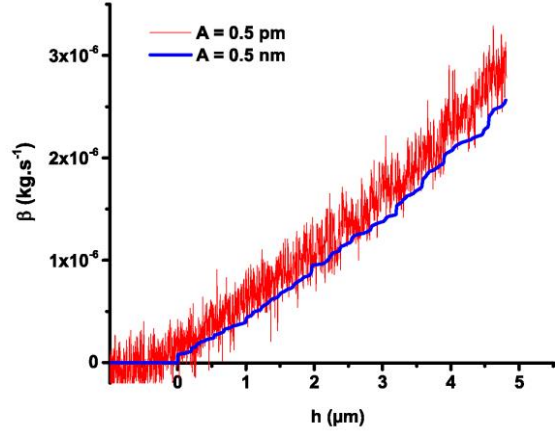

**Figure SM5.** Friction coefficient measured by MEMS-AFM for two oscillation amplitudes 0.5 pm in red and 0.5 nm in blue.

## SM5: Velocity profile and shear stress

The velocity field of a flow induced, in a purely viscous Newtonian liquid, by an infinite cylindrical rod of radius R oscillating along its axis is solution, under the lubrication approximation, of the Stokes equation. In the complex domain and together with a no-slip boundary condition at the rod's surface and a vanishing-speed boundary condition at infinity, it reads:

$$v^*\left(\frac{r}{\delta},\frac{R}{\delta},t\right) = v_0\, e^{-i\omega t} \times f^*\left(\frac{r}{\delta},\frac{R}{\delta}\right) = v_0\, e^{-i\omega t} \times \left(f_1\left(\frac{r}{\delta},\frac{R}{\delta}\right) + i \times f_2\left(\frac{r}{\delta},\frac{R}{\delta}\right)\right)$$

where $\delta = \sqrt{\frac{2\eta}{\rho\omega}}$ is the skin thickness of the corresponding 2D case (of a viscous flow induced by an oscillating plate) and where $f_1$ and $f_2$ are respectively the real and imaginary parts of the function f, which reads:

$$f(u_1, u_2) = \frac{\mathcal{K}_0[(1-i)\, u_1]}{\mathcal{K}_0[(1-i)\, u_2]}$$

with $\mathcal{K}_n$ being the modified Bessel function of the second kind of order *n*.

The real velocity field $v\left(\frac{r}{\delta},\frac{R}{\delta},t\right)$ oscillates thus at the circular frequency $\omega$, in an envelope which will be later on referred as $v_e\left(\frac{r}{\delta},\frac{R}{\delta}\right)$. This envelope corresponds to the maximum of $v\left(\frac{r}{\delta},\frac{R}{\delta},t\right)$ for given $\frac{r}{\delta}$ and $\frac{R}{\delta}$ ratios, that is:

$$v_e\left(\frac{r}{\delta},\frac{R}{\delta}\right) = v_0 \times max|_t \left\{\cos(\omega t) \times f_1\left(\frac{r}{\delta},\frac{R}{\delta}\right) + \sin(\omega t) \times f_2\left(\frac{r}{\delta},\frac{R}{\delta}\right)\right\}$$

thus

$$v_e = v_0 \times \left|\cos\left(\text{atan}\left(\frac{f_2}{f_1}\right)\right) \times f_1 + \sin\left(\text{atan}\left(\frac{f_2}{f_1}\right)\right) \times f_2\right|$$

Similarly, the complex shear stress exerted on the rod's surface reads:

$$\sigma^*\left(\frac{R}{\delta},t\right) = -\eta\ \partial_r v^*|_{r=R} = \sqrt{2} \times \frac{\eta}{\delta} \times e^{-i(\omega t + \frac{\pi}{4})} \times \left(g_1\left(\frac{R}{\delta}\right) + i \times g_2\left(\frac{R}{\delta}\right)\right)$$

where the function g reads:

$$g(u) = \frac{\mathcal{K}_1[(1-i)\,u]}{\mathcal{K}_0[(1-i)\,u]} = g_1(u) + i \times g_2(u)$$

with $g_1$ and $g_2$ being respectively the real and imaginary parts of the function g.

The real shear stress oscillates also at the circular pulsation $\omega$ in an envelope $\sigma_e\left(\frac{R}{\delta}\right)$ where

$$\sigma_e\left[\frac{R}{\delta}\right] = \sqrt{2} \times \frac{\eta}{\delta} \times \left|\cos\left(\text{atan}\left(\frac{g_2}{g_1}\right)\right) \times g_1 + \sin\left(\text{atan}\left(\frac{g_2}{g_1}\right)\right) \times g_2\right|$$

One can thus define the stress enhancement factor $g_e\left(\frac{R}{\delta}\right)$:

$$g_e = \frac{\sigma_e}{\sqrt{2} \times \frac{\eta}{\delta}} = \left|\cos\left(\text{atan}\left(\frac{g_2}{g_1}\right)\right) \times g_1 + \sin\left(\text{atan}\left(\frac{g_2}{g_1}\right)\right) \times g_2\right|$$